\documentclass[twocolumn,showpacs,preprintnumbers,amsmath,amssymb,prl,superscriptaddress]{revtex4}
\usepackage{graphicx}
\usepackage{epstopdf}
\usepackage{dcolumn}
\usepackage{bm}
\usepackage{amsmath}
\usepackage{amssymb}
\usepackage{mathrsfs}
\usepackage{amsfonts}
\usepackage{color}

\usepackage{dsfont}

 \newcommand{\ket}[1]{\ensuremath{|#1\rangle}}
\newcommand{\bra}[1]{\ensuremath{\langle #1|}}
\newcommand{\braket}[1]{\ensuremath{\langle #1 \rangle}}
\newcommand{\op}[1]{\ensuremath{\hat #1}}
\newcommand{\opdag}[1]{\ensuremath{\hat{#1}^\dagger}}

\begin{document}
%%%%%%%%%   TITLE  %%%%%%%%% 
\title{Single Polariton Optomechanics} 

%%%%%%%%% AUTHORS  %%%%%%%%% 
\author{Juan \surname{Restrepo}}
\affiliation{Laboratoire Mat\'eriaux et Ph\'enom\`enes Quantiques, Universit\'e Paris Diderot, CNRS UMR 7162, Sorbonne Paris Cit\'e, 10 rue Alice Domon et L\'eonie Duquet 75013 Paris, France}

\author{Cristiano \surname{Ciuti}}
\affiliation{Laboratoire Mat\'eriaux et Ph\'enom\`enes Quantiques, Universit\'e Paris Diderot, CNRS UMR 7162, Sorbonne Paris Cit\'e, 10 rue Alice Domon et L\'eonie Duquet 75013 Paris, France}

\author{Ivan \surname{Favero}}
\affiliation{Laboratoire Mat\'eriaux et Ph\'enom\`enes Quantiques, Universit\'e Paris Diderot, CNRS UMR 7162, Sorbonne Paris Cit\'e, 10 rue Alice Domon et L\'eonie Duquet 75013 Paris, France}

%%%%%%%%%  DATE  %%%%%%%%% 
\date{\today}

%%% 				Abstract 					%%%

\begin{abstract}
This letter investigates a hybrid quantum system combining cavity quantum electrodynamics and optomechanics. The Hamiltonian problem of a photon mode coupled to a two-level atom via a Jaynes-Cummings coupling and to a mechanical mode via radiation pressure coupling is solved analytically. The atom-cavity polariton number operator commutes with the total Hamiltonian leading to an exact description in terms of tripartite atom-cavity-mechanics polarons. We demonstrate the possibility to obtain cooling of mechanical motion at the single-polariton level and describe the peculiar quantum statistics of phonons in such unconventional regime. 
\end{abstract}


\maketitle

%%%%%%%%%%%%%%%%%%%%%%%%%%%%%%%%%
%%%	  			Introduction					%%
%%%%%%%%%%%%%%%%%%%%%%%%%%%%%%%%%

Cavity Quantum Electrodynamics (QED) experiments have explored light-matter interaction at the quantum level in atomic physics \cite{Haroche06,Kimble98}. Spectacular developments have been achieved as well in circuit QED systems based on superconducting Josephson junctions \cite{Wallraff}. More recently, quantum optomechanical realizations have coupled cavity photons to mesoscopic mechanical resonators \cite{Favero09, Marquardt09,Aspelmeyer13}. In superconducting circuits, strong coupling and control of the mechanical motion at the quantum level have also been demonstrated \cite{Cleland}. Today, the maturity of solid-state quantum devices appears thus promising to bridge QED and optomechanics. The physical interaction at play in QED results in a resonant coupling linear in the photon field operators (Jaynes-Cummings Hamiltonian), while in optomechanics a non-linear radiation pressure term couples two off-resonant photonic and mechanical modes. A rich physics is expected in systems that would merge these distinct physical features.

The basic principle of inserting a two-level artificial atom in an optomechanical setting was discussed in classical terms for fine tuning of dispersive and dissipative optomechanical interactions \cite{Favero08}. The coupling of an optomechanical cavity to an atom motion \cite{Hammerer09} or to collective excitations of an ensemble of atoms \cite{Genes09} was also discussed, resulting in the physical situation of two linearly coupled harmonic oscillators. In that case the anharmonic internal structure of a single atom and its corresponding nonlinear dynamics, a key feature of cavity and circuit QED, is absent. Since optomechanical systems progressively move towards regimes where single photon coupling exceeds dissipation \cite{Rabl11, Nunnenkamp11,Chan11,Ding10,Ding11,Fainstein13} a growing interest is emerging for hybrid systems where artificial atoms, photons and phonons would all be strongly coupled at the quantum level.

In this letter, we investigate the physics of a hybrid quantum system where a cavity photon mode is coupled to an artificial two-level atom and to a mechanical resonator. We describe analytically the polaron excitations of this tripartite system and determine the dynamics in presence of losses and driving. We show atom-assisted cooling of mechanical motion down to the single atom-cavity polariton level and reveal unusual mechanical amplification. Last, we demonstrate the emergence of phonon antibunching in such tripartite quantum systems.

\begin{center}
\begin{figure}[h]
	\includegraphics[scale=1.00]{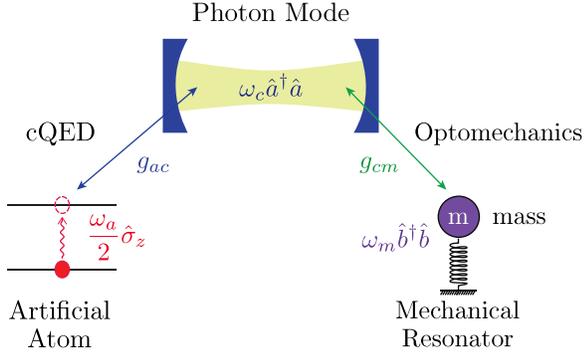}
	\caption{ (Color online) Scheme of the considered hybrid system. A photon confined mode of frequency $\omega_c$ couples both to a two-level system ($\omega_a$ is the transition frequency) and to a mechanical resonator of frequency $\omega_m$. $g_{ac}$ ($g_{cm}$) is the coupling strength of the Jaynes Cummings (radiation pressure) atom-cavity (cavity-mechanics) coupling.}
	\label{fig:scheme}
\end{figure}
\end{center}

%%%%%%%%%%%%%%%%%%%%%%%%%%%%%%%%%
%%%	  		Model of the system					%%%
%%%%%%%%%%%%%%%%%%%%%%%%%%%%%%%%%
As illustrated in Fig.\ref{fig:scheme}, we consider  a joint system where a confined photon mode is coupled both to a two-level artificial atom and to a mechanical resonator. Our system combines the usual Jaynes-Cummings  (JC) coupling of cavity (circuit) QED architectures \cite{Haroche06} and the nonlinear coupling of optomechanics \cite{Law95}. We thus consider the total Hamiltonian ($\hbar=1$):
\begin{equation}
\begin{split}
\op{H}_{tot}=&\omega_c \opdag{a}\op{a}+\frac{\omega_a}{2}\op{\sigma}_z+ig_{ac} \left(\op{\sigma}_+\op{a}-\op{\sigma}_- \opdag{a} \right) \\
& +\omega_m \opdag{b}\op{b}-g_{cm}\opdag{a}\op{a} \left( \op{b}+\opdag{b} \right),\end{split}
\label{eq:FullHamiltonianSystem}
\end{equation}
where $\op{\sigma}_{x,y,z}$ are Pauli matrices for the two-level system  ($\op{\sigma}_{\pm}$ being the  ladder operators), $\op{a}$ ($\op{b}$) is the annihilation operator for the photon (mechanical) mode of frequency $\omega_c$ ($\omega_m$). $g_{ac}$ and $g_{cm}$ are the atom-cavity  and optomechanical coupling strengths. Replacing the radiation pressure coupling with a direct coupling between the atom and the mechanical oscillator of the form $\propto \op{\sigma}_z (\op{b}+\opdag{b})$ yields qualitatively a similar physics and this case will hence not be considered further below.
%%%%%%%%%%%%%%%%%%%%
%%% 		 Figure1		        %%%
%%%                end fig1 		       %%%
%%%%%%%%%%%%%%%%%%%%
The system evolves in a discrete but infinite Hilbert space in which the uncoupled ($g_{ac}=g_{cm}=0$) eigenvectors are labeled as $\ket{e,g}\otimes \ket{k} \otimes \ket{l}$:  $\ket{g}$ ($\ket{e}$) describes the ground (excited) state of the two-level system, $k\in \mathbb{N}$ the Fock state with $k$ photons in the cavity and $l\in \mathbb{N}$ the state with $l$ phonons in the mechanical resonator. For $g_{ac} \neq 0$ and $g_{cm} = 0$, one has 
the standard configuration of cavity (circuit) QED, where atom-cavity polaritons are the new eigenmodes of the system. For  $g_{cm} \neq 0$, there are new eigenmodes for the hybrid tripartite system, which we will call atom-cavity-mechanics polarons.

\begin{center}
\begin{figure}[h]
	\includegraphics[scale=0.75]{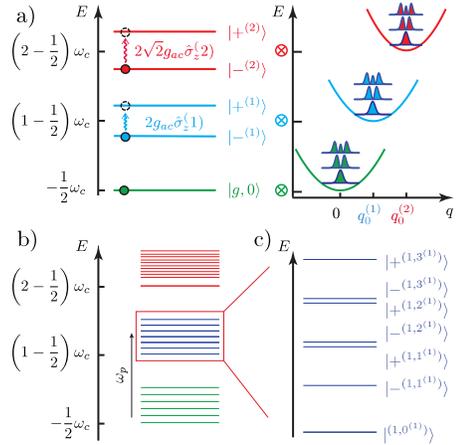}
	\caption{Structure of the polaron eigenstates for the atom-cavity-mechanics tripartite system. a) Each $n$-polariton subspace (left) couples independently to states of the mechanical resonator with different displacements (right). b) Energies spectrum for $n\leq2$. c) Zoom on the energy levels in the ${\mathcal H}_1$ subspace (1-polariton states).}	
	\label{fig:diagonalisation}
\end{figure}
\end{center}

A key feature of the above Hamiltonian is the conservation of the polariton number  $\op{N}_{polariton}=\opdag{a}\op{a}+\op{\sigma}_+ \op{\sigma}_-$. The problem can thus be diagonalized in each subspace ${\mathcal H}_n$ containing exactly $n$ polaritons. In the resonant case ($\omega_a=\omega_c$) the polariton-phonon basis $\{\ket{\pm^{(n)}}\otimes\ket{l} \}$ diagonalizes the JC part of the Hamiltonian $\op{H}_{JC}=\omega_c \opdag{a}\op{a}+\frac{\omega_a}{2}\op{\sigma}_z+ig_{ac} \left(\op{\sigma}_+\op{a}-\op{\sigma}_- \opdag{a} \right)$, where $\ket{\pm^{(n)}}=\frac{1}{\sqrt{2}} \left( \ket{g,k=n}\pm i \ket{e,k=n-1} \right)$ are the $n$-polariton eigenvectors of the JC ladder.
Namely, we have
$\op{H}_{JC} \ket{\pm^{(n)}}=\omega_{\pm}^{(n)} \ket{\pm^{(n)}}$ with $ \omega_{\pm}^{(n)} = \left( (n-1/2)\omega_c \pm \frac{\Omega^{(n)}}{2} \right)$ and 
$\Omega^{(n)}=2\sqrt{n}g_{ac}$.  Being the system total Hamiltonian  block diagonal in this new basis, the mechanical resonator couples independently to each polaritonic subspace associated to ${\mathcal H}_n$, as illustrated in Fig.\ref{fig:diagonalisation}.a. Hence, we can express it as follows:
\begin{equation}
	\begin{split}
		\op{H}_{tot}=& \sum_{n\in \mathbb{N}} \Bigg\{ (n-1/2)\omega_c \mathds{1}^{(n)} + \frac{\Omega^{(n)}}{2} \op{\sigma}^{(n)}_z  \\
		                     ~&  -g_{cm} \left(\frac{1}{2}\op{\sigma}^{(n)}_x  +  (n-\frac{1}{2}) \mathds{1}^{(n)}\right) \left(\op{b}+\opdag{b} \right) \Bigg \}+ \omega_m \opdag{b}\op{b}
	\end{split}
	\label{eq:JcOptomechanics}
\end{equation}

where  $\mathds{1}^{(n)}$ is the identity in the ${\mathcal H}_n$ Hilbert subspace, $\op{\sigma}_{x,z}^{(n)}$ are Pauli matrices acting on the polaritonic doublet $\ket{\pm^{(n)}}$. The radiation pressure coupling has two consequences. First, an effective  coupling appears between each of the two cavity polariton states $\ket{\pm^{(n)}}$ and the mechanical mode. Second, each of the two states contains on average $n-1/2$ cavity photons that displace statically the equilibrium position of the mechanical resonator.

In each ${\mathcal H}_n$ we can absorb the static displacement of the mechanical resonator by introducing the new displaced operator $\op{b}_{n}$ and make a rotating wave approximation to obtain the following Hamiltonian in ${\mathcal H}_n$:

\begin{equation}
	\begin{split}
		\op{H}^{(n)}=&\frac{\Omega^{(n)}}{2}\op{\sigma}_z^{(n)}+\omega_m \opdag{b}_{n} \op{b}_{n}-\frac{g_{cm}}{2} \left(\op{\sigma}_-^{(n)}\opdag{b}_{n}+\op{\sigma}_+^{(n)}\op{b}_{n} \right)\\
				    &-g_{cm}\frac{\sqrt{2}}{2}q_0^{(n)}\op{\sigma}_x^{(n)}+\left( \omega_0^{(n)}-\frac{\omega_m}{2}q_0^{(n)^2} \right) \mathds{1}^{(n)}
	\end{split}
	\label{eq:HnShifted}
\end{equation}	

where $q_0^{(n)}=\sqrt{2}g_{cm}(n-1/2)/\omega_m$ is the displaced mechanical equilibrium position. The term proportional to $\op{\sigma}_x^{(n)}$ in Eq.\ref{eq:HnShifted} will be neglected hereafter in the limit $g_{cm}<<\omega_m$. Its perturbative effect could be accounted for in an exact treatment of the Rabi model \cite{Braak11}  but at the expense of losing explicit expressions for the eigenstates.  Apart from the last term which is a constant energy shift, we have an effective Jaynes-Cummings-like Hamiltonian, which can be diagonalized by the states $\ket{\pm^{(n,m^{(n)})}}$, where $m^{(n)}$ is the polaron number of the tripartite system and $n$ is the polariton number previously defined. The energy spectrum of $\op{H}_n$ is given by the expression:

\begin{equation}
	\op{H}^{(n)}\ket{\pm^{(n,m^{(n)})}}=\omega_0^{(n)}-\frac{\omega_m}{2}q_0^{(n)^2} +\left(m-\frac{1}{2} \right) \omega_m \pm \nu^{(n,m)}
\end{equation}
where,
\begin{equation}
	\nu^{(n,m)}=\sqrt{\left( \frac{\Omega^{(n)}-\omega_m}{2}\right)^2+\frac{m^{(n)}}{4}g_{cm}^2}
\end{equation}		

%%%%%%%%%%%%%%%%%%%%	
%%% 		 Figure2		        %%%
%%%                end fig2		       %%%
%%%%%%%%%%%%%%%%%%%%
	
Fig.\ref{fig:diagonalisation}b) shows the structure of the energy spectrum for $n = 0, 1, 2$. For each $n$-polariton state, a multiplet arises from the mechanical resonator Fock space. As the mean number of photons $n-1/2$ increases, the anharmonicity of the system becomes more pronounced within each multiplet. For illustrative purposes Fig.\ref{fig:diagonalisation}c depicts in detail the anharmonic spectrum of the $n=1$ multiplet.

%%%%%%%%%%%%%%%%%%%%%%%%%%%%%%%%%%
%%%	  		Dynamics of the open system			 %%%
%%%%%%%%%%%%%%%%%%%%%%%%%%%%%%%%%%

Having understood the nature of the dressed polaron states, we can consider the dynamics of the open system in presence of losses and when coupled to a bath at temperature $T$. Within a Lindblad approach for the dissipation, the density matrix $\op{\rho}$ of the system follows the master equation

\begin{equation}
	\begin{split}
	\frac{d \op{\rho}(t)}{dt}=&-i[\op{H}_{tot}+\op{V}_{p}(t),\op{\rho}] +\gamma_{ac} L[\op{a}] \op{\rho}+\gamma_{ac} L[\op{\sigma}_-] \op{\rho}\\
					   & +n_{th} \gamma_m L[\opdag{b}] \op{\rho}+(n_{th}+1) \gamma_m L[\op{b}]\op{\rho}
	\end{split}
	\label{eq:ME}
\end{equation}		

where $\gamma_m$ ($\gamma_{ac}$) is the phonon (polariton) loss rate, $n_{th}$ is the thermal mean phononic occupancy, $\op{V}_{p}(t)=iF_p (\opdag{a} e^{i \omega_p t}-\op{a} e^{-i \omega_p t})$ is a coherent pump term with frequency $\omega_p$ and $L[\op{o}]\op{\rho}=\op{o} \op{\rho} \opdag{o}-1/2(\opdag{o}\op{o}\op{\rho}+\op{\rho}\opdag{o}\op{o}) $ for any given jump operator $\op{o}$.

\begin{center}
\begin{figure}[h]
	\includegraphics[scale=1]{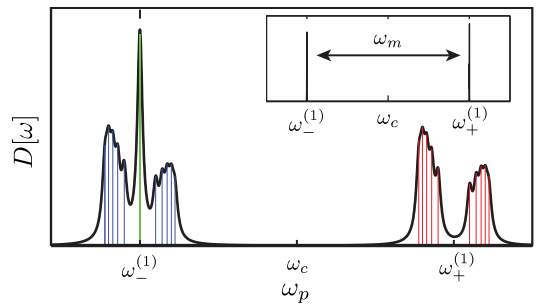}
	\caption{ Optical joint spectral density of polaronic states describing transitions between the states with $0$ and $1$ polaritons. Main panel: $g_{cm}/\omega_m=10^{-1} $,  $Q_{ac}=10^4$. Insight:  $g_{cm}/\omega_m=10^{-3}$, $Q_{ac}=10^{6}$. For clarity, we only present transitions between states with $m^{(n)}\leq5$ polarons. Blue, green and red peaks (color online) represent transitions reducing, conserving and increasing the number of phonons respectively.}	
	\label{fig:spectralJointDensity}
\end{figure}
\end{center}

For strong enough light-matter coupling the anharmonicity of the polaritonic  energies is known to give rise to photon blockade effects \cite{Werner99}. In the following we only consider a moderate pumping regime in which one cavity photon and hence one single polariton is excited at most. This allows to  understand the physics in terms of the subspaces ${\mathcal H}_n$ with $n\leq 1$ and hence to consider only transitions within the set of states $\{ \ket{\pm^{(n,m^{(n)})}} \} _{n \in [0,1]}$ 
(note: the numerical solutions actually include states with higher number $n$, the validity of such approximation has been carefully checked). In the figures of the manuscript,  unless otherwise stated, we will always consider the following parameters: $\omega_c/\omega_m=10^2$, $\omega_a=\omega_c$, $g_{ac}/\omega_m=1/2$, $g_{cm}/\omega_m=10^{-1} $, $Q_m=\omega_m/\gamma_m=10 ^4$, $Q_{ac}=\omega_{a,c}/\gamma_{ac}=10 ^4$, $F_p/\gamma_{ac}=1$ and $n_{th}=3.45$.\\
 We initially prepare the system in the state $\ket{g,k=0}\bra{g,k=0}\otimes \op{\rho}_m$ (no photons in the cavity, atom in the ground state) and pump it  with photons having an energy close to the energy difference between the 1-polariton and 0-polariton states,  as can be seen in Fig.\ref{fig:diagonalisation}.b. In the single polariton regime considered here, the polaritonic splitting $\Omega^{(1)}$ is tuned  to about the mechanical frequency $\omega_m$ in order to make resonant the processes in which a phonon is annihilated (created)  thus giving rise to cooling (amplification) . These processes can be selected optically by a proper tuning of the pump frequency $\omega_p$.  Fig.\ref{fig:spectralJointDensity} presents the joint spectral density of polaronic states defined as follows:

 \begin{equation}
	D[\omega]=\sum_{\substack{s',s=\pm \\m',m \in \mathbb{N}}} |\bra{s'^{(1,m')}}\op{V}_p \ket{s^{(0,m)}}|^2 \delta [ \omega-(\omega_	{s'^{(1,m')}}-\omega_{s^{(0,m)}})  ].
	\label{eq:oSPJD}
\end{equation}

%%%%%%%%%%%%%%%%%%%%
%%% 		 Figure3		        %%%
%%%                end fig3		       %%%
%%%%%%%%%%%%%%%%%%%%

For clarity in Fig.\ref{fig:spectralJointDensity} we have convoluted the spectral joint density with a lorentzian of width $\gamma_{ac}$ for both cases. $D[\omega]$ shows a series of resonances corresponding to transitions where the number of phonons is either conserved, reduced  or increased. The main panel shows $D[\omega]$ for $g_{cm}/\omega_m=10^{-1}$ displaying two spectral structures centered on the lower (upper) polariton energy $\omega_-^{(1)}$ ($\omega_+^{(1)}$). Each structure is itself split into a doublet with a splitting $\sim g_{cm}$. The inset of Fig.\ref{fig:spectralJointDensity} shows $D[\omega]$  for a weaker optomechanical coupling $g_{cm}/\omega_m=10^{-3}$. In this case the polaronic fine structure splitting $\sim g_{cm}$ is no longer visible at a scale $\sim \omega_m$ and $D[\omega]$ presents only two resonances at $\omega_{\pm}^{(1)}$. The resonances around $\omega_-^{(1)}$ correspond to transitions for which the number of phonons decreases (blue peaks) or is conserved (green peak), those around  $\omega_+^{(1)}$ correspond to processes increasing the number of phonons (red peaks). 

%%%%%%%%%%%%%%%%%%%%%%%%%%%%%%%%%%%%%%%%%%%%%%
%%%	  		Single polariton cooling of mechanical motion					%%%
%%%%%%%%%%%%%%%%%%%%%%%%%%%%%%%%%%%%%%%%%%%%%%

\begin{center}
\begin{figure}[h!]
	\includegraphics[scale=1.00]{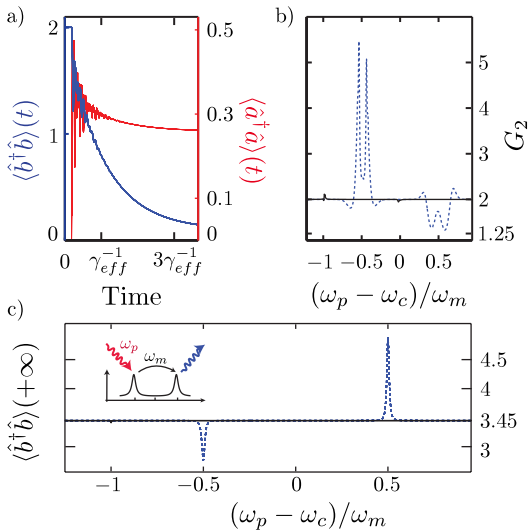}
	\caption{(color online) Atom-assisted optomechanical cooling. a) Time evolution of the number of photons (red) and phonons (blue) for an initial mechanical state $\ket{l=2}$. b) Phonon second-order correlation function $G_2$ of the stationary state. c) Stationary number of phonons as a function of $\omega_p$ for  $Q_m=10^{6}$, $Q_{ac}=10^{6}$, $g_{cm}/\omega_m=10^{-3}$,  $F_p/\gamma_{ac}= 100$. In b) and c) the dashed blue lines represent the hybrid QED-optomechanics case with an atom ($g_{ac}\neq0)$ while the black solid lines correspond to the usual atomless scenario($g_{ac}=0$). The inset of c) depicts schematically the doubly-resonant polariton cooling of motion.}	
	\label{fig:cooling}
\end{figure}
\end{center}

 Fig.\ref{fig:cooling}a) presents the dynamical behavior of the mechanical mode coupled to the polaritonic atom-cavity system under optical pumping close to $\omega_-^{(1)}$ (cooling condition), for the set of parameters of Fig.\ref{fig:spectralJointDensity} main panel. It presents the number of phonons and photons as a function of time  (pump switched on abruptly at $t>0$, inducing an early transient regime). As the number of photons in the cavity approaches a stationary state, the number of phonons steadily decreases with an effective dissipation constant $\gamma_{eff} \simeq 18 \gamma_m$ towards an asymptotic value $n_{min} \sim 1/10$. The dependence of $\gamma_{eff}$ and $n_{min}$ on the pump  frequency $\omega_p$ shows a good fit with the transitions described by $D[\omega]$ (not shown here). By numerically finding the stationary solution of Eq.\ref{eq:ME} it is possible to study the system statistics in the stationary regime ($t \to +\infty$). Fig.\ref{fig:cooling}.b presents the stationary second order autocorrelation function $G_2=\braket{\opdag{b}\opdag{b}\op{b}\op{b}}/\braket{\opdag{b}\op{b}}^2$ as a function of the pump  frequency.  When the transitions around the lower polariton energy are excited (here $\omega_-^{(1)}=\omega_c- \omega_m/2$) the mean phonon occupancy is reduced as previously mentioned and the statistics of the mechanical oscillator is drastically changed leading to strong  phonon bunching $G_2\sim 5$. The solid line represents the values of $G_2$ for the equivalent atomless system ($g_{ac}=0$). The changes on $G_2$ in this scenario are much weaker ($\sim10^{-2}$) showing that the presence of the two-level atom is crucial for entering this strong phonon bunching regime.
Fig.\ref{fig:cooling}.c reports the stationary number of phonons as a function of $\omega_p$ for the set of parameters of the inset of Fig.\ref{fig:spectralJointDensity}, showing both cooling and amplification as the optical pump resonates with the polaritonic levels. The dashed line represents the evolution of $\braket{\opdag{b}\op{b}}(+\infty)$ for the corresponding atomless scenario ($g_{ac}=0$), the other parameters remaining the same. In this case the cooling mechanism is hindered by the non resonant pumping, and no phonon population change is in practice visible. In the case considered in Fig.\ref{fig:cooling}, the insertion of the two-level atom in the optomechanical cavity strongly boosts the cooling of mechanical motion through a doubly resonant process. 

%%%%%%%%%%%%%%%%%%%%
%%% 		 Figure4		        %%%
%%%                end fig4		  %%%
%%%%%%%%%%%%%%%%%%%%

%%%%%%%%%%%%%%%%%%%%%%%%%%%%%%%%%
%%% 			Peculiar phonon statistics 			%%%
%%%%%%%%%%%%%%%%%%%%%%%%%%%%%%%%%		

We now move to situations where phonon amplification or emission can occur under pumping of the hybrid atom-optomechanical system. Fig.5 exhibits the phonon statistics (with a zero-temperature bath) for different polariton pumping regimes. Fig.5a) and b) correspond to a coherent pump whose frequency has been set close to the upper-polariton energy. This pump detuning leads to the appearance of non-classical statistics for the mechanical motion as shown in Fig.5b) where the mechanical Wigner function acquires negative values (represented by a black ring in the density plot). Describing analytically this behavior appears quite involved. We checked however that the presence of the atom is mandatory to obtain this non-classicity for the considered set of parameters (see supplements). We also checked that a finite temperature for the phonon bath progressively destroys the negativity of the Wigner function. In some cases, interesting analogies can be drawn between this polariton-assisted amplification of motion and the situation of a single-atom laser \cite{delValle11}. These analogies are addressed in the supplements.

If we now consider the case of an incoherent pumping on the 1-polariton states (described by adding $F_{inc}(L[\ket{+^{(1)}}\bra{g,0}]\op{\rho}+L[\ket{-^{(1)}}\bra{g,0}]\op{\rho})$ to Eq.\ref{eq:ME})
the phonon statistics can be described through an analytical approach  \cite{delValle11,supplementary}. As shown in Fig.5c) the incoherent pump populates the excited polaritonic states which, in the eigenbasis of the Hamiltonian, leads to excitation of polarons and then emission of phonons. Fig.5d) shows that for weak incoherent pumping the anharmonicity of the polaronic states yields sub-poissonian statistics ($G_2\ll1$) for the emitted phonons. The analytical results (circles) are in excellent agreement with the numerical results.
%%%%%%%%%%%%%%%%%%%%
%%% 		 Figure5		        %%%
\begin{center}
\begin{figure}[h!]
	\includegraphics[scale=1.00]{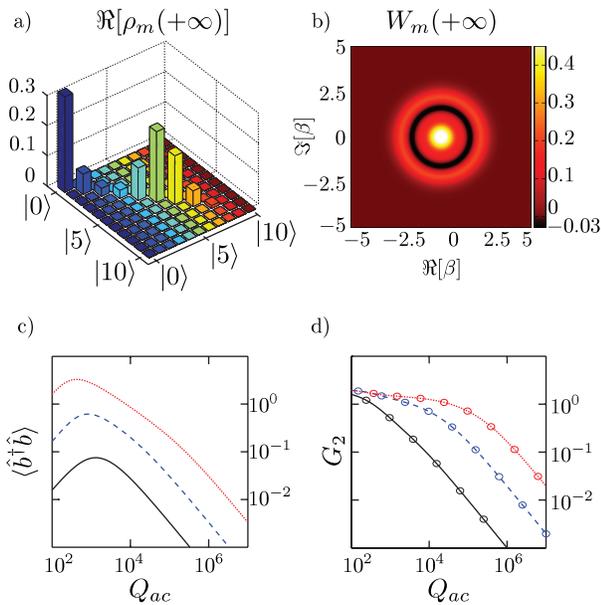}
	\caption{(color online) Steady-state mechanical resonator statistics (bath at zero temperature). a) Real part of the mechanical reduced density matrix and b) corresponding Wigner function for an optical coherent pump tuned to $(\omega_p-\omega_+^{(1)})/\omega_m=10^{-1}$. c) Phonon occupation number and d)  second-order correlation function $G_2$ as a function of the polariton quality factor $Q_{ac}$. The incoherent pump rate is set to $F_{inc}=\gamma_{ac}$. Black-solid, blue-dashed and red-dotted lines correspond to numerical results for $Q_m=10^1,10^2$ and $10^3$ respectively. The analytical solutions of the master equation are represented by circles.}
	\label{fig:amplification}
\end{figure}
\end{center}
%%%                end fig5		  %%%
%%%%%%%%%%%%%%%%%%%%

%%%%%%%%%%%%%%%%%%%%%
%%% 		Conclusion 		%%%
In conclusion, we have proposed a hybrid system where a mechanical resonator is coupled to a cavity (circuit) QED system embedding a single artificial two-level atom. The physics of the system can be described in terms of atom-cavity-mechanics polarons. In presence of losses, we have focused on the single polariton optomechanics and shown that an artificial two-level atom can enhance single-photon cooling by orders of magnitude and lead to strong bunching of phonons. When the bath of the system is at very low temperatures, the hybrid configuration leads to non-classical statistics of the mechanical motion leading to negative values of its Wigner function and emission of single phonons with strong anti-bunching. These concepts could be tested on a large set of experimental platforms spanning from a real single atom trapped in a macroscopic optomechanical Fabry-Perot cavity resonator, to diamond resonators embedding  NV centers, defect centers in silica toroids \cite{Ramos13}, or semiconductor optomechanical systems with built-in artificial atoms. Of peculiar interest are miniature Gallium-Arsenide optomechanical resonators combining strong optomechanical \cite{Ding10,Ding11} with cavity QED couplings \cite{Peter05}, and superconducting systems where strong coupling circuit QED and quantum control of GHz mechanical motion have been already demonstrated \cite{Cleland}. These hybrid atom-cavity-mechanics platforms are not far from existing in laboratories and will allow to transfer mature concepts of cavity (circuit) QED to the realm of mechanical systems. 

%%%%%%%%%%%%%%%%%%%%%%%%%%%
%%%	     		 Bibliography 		 	   %%%
%%%%%%%%%%%%%%%%%%%%%%%%%%%

\end{document}